\documentclass{article}

\usepackage{harvard}
\usepackage{graphics}
\usepackage{amssymb}

\def\url#1{{\ttfamily\def\/{/\discretionary{}{}{}}#1}}

\def\ea{{\it et al.\,}}
\def\be{\begin{equation}}
\def\ee{\end{equation}}
\def\eg{{\it e.g., \,}}

\def\ie{{\it i.e., \,}}
\def\apj{{\it Astroph.J. \,}}
\def\ao{{\it Applied Optics \,}}
\def\address{{\it  \,}}

\begin{document}

\title{Absolute calibration and beam reconstruction of MITO (a ground-based instrument in the millimetric region).}
\maketitle
\author{G. Savini$^{a}$}
\author{A. Orlando$^{a}$}
\author{E.S. Battistelli$^a$}
\author{M. De Petris$^a$}
\author{L. Lamagna$^a$}
\author{G. Luzzi$^a$}
\author{E. Palladino$^a$}

\address{$^a$  Dipartimento di Fisica, Universit\'a "La Sapienza", P.le A.Moro 2, I-00185
Roma,Italy}

\begin{abstract}
An efficient sky data reconstruction derives from a precise
characterization of the observing instrument. Here we describe the
reconstruction of performances of a single-pixel 4-band photometer
installed at MITO (Millimeter and Infrared Testagrigia
Observatory) focal plane. The strategy of differential sky
observations at millimeter wavelengths, by scanning the field of
view at constant elevation wobbling the subreflector, induces a
good knowledge of beam profile and beam-throw amplitude, allowing
efficient data recovery. The problems that arise estimating the
detectors throughput by drift scanning on planets are shown.
Atmospheric transmission, monitored by skydip technique, is
considered for deriving final responsivities for the 4 channels
using planets as primary calibrators.
\end{abstract}

\section{Introduction}
At millimeter wavelengths ground based cosmological observations
are mainly constrained by atmospheric transmission and its time
dependent emission as huge noise source. The largely employed
observational technique, among collectors of submm/mm radiation,
is differential measurements and consequent synchronous
demodulation, $i.e.$ lock-in amplifiers technique. At MITO, in
order to reduce the bulk of atmospheric emission, sky differential
observations are obtained by a wobbling sub-reflector in a
Cassegrain 2.6-m telescope. Modulation in the sky is usually
performed as a 3-field square-wave function as to remove linear
gradients of atmosphere in every direction. In the data recovery
exact beam-throw amplitude and beam-shape have to be recovered for
successive data analysis. This can be obtained, as we will show,
by simulating the effect of systematics on the time ordered data.
Different methods for beam shape extraction are then explained,
and differences between the effects on calibration on point-source
and extended beam-size objects outlined. Revised responsivity
values obtained with the considered Jupiter temperature models are
reported in section 4. In the last section, geometrical
consequences on galaxy-cluster SZ-effect measurements are
analyzed.
\section{Data acquisition system.}
Ground based telescopes observing a specific object in the sky can
operate mostly in two different ways. One is chasing the object
while moving in the sky, and the other is drift-scanning.
Microphonics and side-lobe peak-ups prevent the possibility of
tracking the source by moving the telescope: all our observations
have been carried out in the so-called drift-scan mode. That is,
the telescope is fixed, pointing a sky region which will be
crossed by the source due to the natural Earth's rotation. Since
we know approximately the shape of our source we may predict the
time-shape of the signal and only its peak amplitude remains a
free parameter. In the present focal plane 4-bands single pixel
photometer (FotoMITO), the voltage-signals of the bolometric
detectors are filtered by commercial lock-in amplifiers (SR850).
The use of lock-in amplifiers will be upgraded with a
software-based acquisition system, allowing a great variety of
offline demodulation techniques, in the new detector array
instrumentation, MAD (Lamagna et al. 2002).

\subsection{Retrieval of exact modulation amplitude.}

Modulation of the secondary mirror is controlled by an
electro-mechanical system with shakers in a push-pull
configuration (Mainella et al. 1996), its position is read by a
Linear Variable Differential Transformer (LVDT). Original optical
design should allow us to retrieve information on sky modulation,
the amplitude of which can be trimmed remotely. The exact
modulation amplitude is checked by drift-scanning objects that can
be treated as point-sources in respect to beam-size (like a planet
$\sim 10'' \div 40''$ in respect to $\sim 15^{\prime}\div
17^{\prime}$ FWHM of f.o.v.). If we assume a sampling rate of 1
Hz, then by knowing the distance between the two minima of the
drift-scan, by adjusting for drift-modulation angle and knowing
the speed of the object by its celestial declination, we can
measure with good approximation the beam-throw of our telescope.
Assumptions are that the shape of the beam is the same during
modulation (primary goal during mirror design) (De Petris et al.
1996). In particular, this assumption has been verified by
comparing the beam shape obtained with different methods
(see.\ref{sec:beam}).
 Precision of this measurement can be mined by
atmospheric fluctuations (greatest noise source) in the proximity
of the minima of the drift-scans (when a culminating source is in
the center of the modulated beam). A solution is to find a
polynomial best-fit around each minimum to obtain a reasonable
identification of relative minima. If acquisition was achieved
without hardware or software lock-in procedures, setting aside
noise problems, and with beam-throws considerably greater than the
width of the beam at -30dB so to avoid contamination and averaging
as we do on 3 modulation periods, we would reasonably have 3
distinct beam-shapes (for a culminating source) separated by a
small baseline. As beam-throw decreases contamination sets in and
the three shapes start to merge.

\begin{figure}[htbp]
    \begin{center}
       \resizebox{!}{\textwidth}{\includegraphics*
            [28mm,100mm][200mm,260mm]{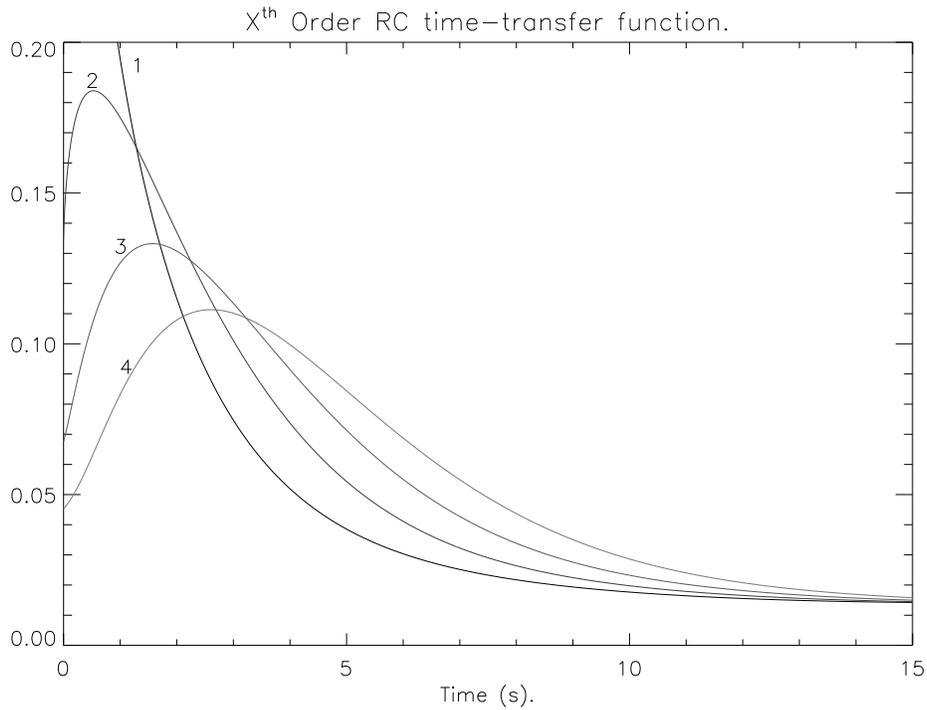}}
        \caption{First to fourth order time
impulse response function of the Lock-in low-pass integration
circuit.} \label{fig:ttf}
    \end{center}
\end{figure}

Moreover Lock-in techniques introduce time-integration due to
their internal filter to extract the locked frequency with a steep
cut. In our case (hardware and software) a $4^{th}$ order RC
circuit filter has been employed (fig.\ref{fig:ttf}). Setting the
time-constant of the lock-in determines the impulse-response that
is applied to acquired data in the time domain. This introduces a
progressive time-shift that becomes constant after a certain
number ($10 \div 20$) of time-constants. If base-lines
(acquisition of atmospheric signal before and after the source
enters the modulating beam) are sufficiently long, the shift sets
in and no correction for time separation of minima is needed. If
not, it would be necessary to "de-convolve" the signal from the
time-impulse, but this can be done only on the final averaged data
points. To do so we sample the time-domain transfer function at
the same rate of our final data (i.e. 1 per second) in a vector $K
=\{ k_i \}$. In this way the data set $D =\{ d_i \}$ can be viewed
as the result of the following operation on the true "would-be"
data $T=\{ t_i \} $:

\begin{equation}
D = \frac{(K\circ T)}{A} \ \ \ \ \  \bigl(\mbox{ with }
A=\sum_{i=0}^{n-1} k_i \bigr)
\end{equation}

In this way if we make the non restrictive hypothesis that
non-acquired data before the beginning of our drift (like an
extended base-line) is zero, or in the case of a constant baseline
with a non-zero signal, equal to that offset (we consider a zero
baseline for calculations), we obtain the recursive process:

\begin{eqnarray}
\lefteqn
 \ d_0 & = & t_0k_0/A ,\nonumber \\ d_1 & = & (t_1k_0 + t_0k_1)/A ,\nonumber \\ d_i & = & \sum
_{j=0}^i (t_{i-j}k_j)/A \ \ \ \mbox{for}\ i<n ,\nonumber  \\ d_i &
= & \sum _{j=0}^{n-1} (t_{i-j}k_j)/A \ \ \ \mbox{for}\ i\ge n
\end{eqnarray}

that under the cited assumptions can be solved backwards in

\begin{eqnarray}
\lefteqn
 \ t_0 & = & Ad_0/k_0 ,\nonumber \\ t_1 & = & (Ad_1- t_0k_1)/k_0 ,\nonumber \\
 t_i & = & \bigl(Ad_i - \sum _{j=1}^{i} (t_{i-j}k_j)\bigr)/k_0 \ \ \ \mbox{for}\ i<n ,\nonumber  \\ t_i &
= & \bigl(Ad_i - \sum _{j=1}^{n-1} (t_{i-j}k_j)\bigr)/k_0 \ \ \
\mbox{for}\ i\ge n
\end{eqnarray}

The obtained vector (fig.2) will be purged from the increasing
time-shift that can distort the first minima, modulation
beam-throw can be thus obtained by estimating the distance between
the two minima fits, $\Delta _m(s)$, and converted by the
following:

\begin{equation}
Beamthrow(')=\frac{\Delta _m(s)}{60\cdot\cos
\alpha}(15''\cdot\cos\delta)
\end{equation}
\begin{equation}
\tan \alpha = \frac{sin H}{\cos \delta \tan \phi-\sin \delta \cos
H}
\end{equation}

where $15'' \cos \delta $ is the true speed of the source in the
sky and $1/\cos \alpha $ accounts for the projection of the minima
position due to the parallactic angle between source passage and
constant elevation modulation as in fig.3 ($\delta$ is the source
declination, $\phi$ is the geographic latitude and $H$ is the Hour
angle, $H=LST-\alpha$, right ascension subtracted from the Local
Sidereal Time ).
 In the case of equatorial modulation such a correction is not
necessary.

\begin{figure}[htbp]
    \begin{center}
        {\resizebox{!}{\textwidth}{\includegraphics*
            [28mm,80mm][200mm,260mm]{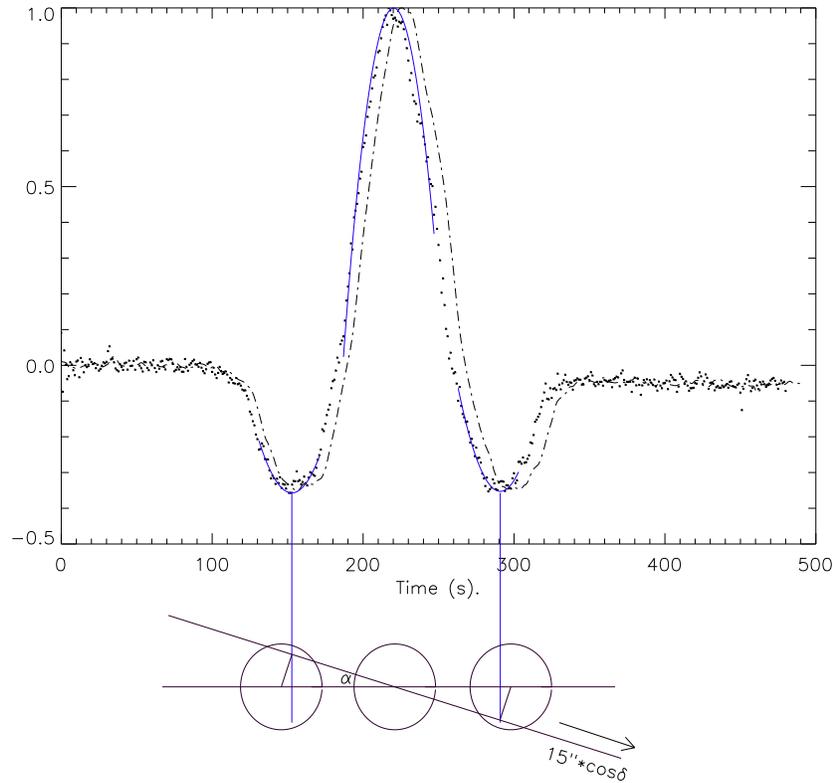}}}
        \caption{Demodulated drift scan of Jupiter
(dots) and acquired drift-scan (dot-dash), with fitted minima and
maximum (short continuous). Increasing delay between the two are
barely visible. Correction for declination and non-transit passage
of source are conceptually drawn under the plots.
 \label{fig:drift}}
    \end{center}
\end{figure}

This procedure has been applied on all planet drift-scans with a
moderate $\alpha$ angle (\ie close to transit) so to have
pronounced minima well above noise level. The beam-throw of our
telescope has been calibrated to a value of $(22.1' \pm 0.4')/Vpp$
of the transducer. Stability of modulation angles during the drift
scans is higher than 0.3\%.

\subsection{Correction for atmospheric absorption.}

In order to correct for atmospheric absorption we measure zenithal
channel transmissions ($\tau_{0}$) performing several skydips
(chopping at focal plane between atmospheric signal and a
Black-Body reference\footnote{Eccosorb AN-72} at telescope
temperature) during each night and correcting for a secant-law at
angles different from zenith. Altitude angles are stepped to match
tenths of air-masses ($90°$ for $1.0$, $65.24°$ for $1.1$, ecc...
). Skydips data present themselves as continuous acquisition of an
offset that varies at the change of telescope elevation. Averages
of each plateau continuous acquisition are subsequently fitted
with a $1/\cos h$ law. As expected there is good agreement with
air-mass dependence and we obtain transmission values for zenithal
angle $z$:

\be
T=\biggl(1-\frac{\rho}{\cos (90°-h)}\biggr) \ee

where $1-\rho $ is zenithal transmission, and $h$ is telescope
elevation.
 Typical atmospheric transmission values for the four channels are listed in table
 \ref{tab:1}.

\begin{table}
\begin{center}
\caption{Values for zenithal transmission of the 4 channels on
jupiter drift night. \label{tab:1}}
%\resizebox{10cm}{10cm}
{\begin{tabular}{cc}
  Channel & $\tau_0$ \\ \hline
  $1 (143 GHz)$ & $0.853$ \\
  $2 (212 GHz)$ & $0.866$ \\
  $3 (271 GHz)$ & $0.812$ \\
  $4 (353 GHz)$ & $0.651$ \\ \hline
\end{tabular}}
\end{center}
\end{table}

\section{Beam shape reconstruction.}\label{sec:beam}

Having a correct knowledge of the modulation amplitude, we now
focus on the exact beam-shape of our instrument. By technical
design, we employ a compact Cassegrain configuration with a 8
meter focal length and 40mm of correct focal plane we expect by
design and ray-tracing a FWHM beam of 17 arcminutes. Yet, it is
necessary to have an experimental evidence of the beam-shape of
photometer+telescope (the beam shape of the photometer alone has
been measured in laboratory). An exact reconstruction of the beam
shape allows us to reduce the contribution to calibration
uncertainty regarding instrumental errors (for planet temperature
see \ref{sec:planets}).

The responsivity of the different spectral channels, when
measuring CMB anisotropies, is defined by eq.\ref{form:resp}. The
two major uncertainties in this quantity are the planet
temperature considered as the integrated black-body temperature
over the frequency bands of the instrument, and the $\Omega
_{beam}$ of the telescope, that is exactly expressed by
eq.\ref{form:AR}

\begin{equation}\label{form:resp}
  R(\mu K/nV)= \frac{T_{CMB}}{S_{planet}}\frac{\Omega _{planet}}{\Omega _{beam}}\frac{\int BB(T_{planet},\nu)f(\nu)d\nu}{\int BB(T_{CMB},\nu)\frac{xe^{x}}{e^{x}-1}f(\nu)d\nu}
\end{equation}
\begin{equation}\label{form:AR}
  \Omega _{beam}= \int \int \sin \theta \cos \theta \cdot AR(\theta ,\phi
  )d\theta d\phi
\end{equation}

where $BB(T,\nu )$ is the blackbody emission (brightness), and
$AR(\theta ,\phi)$ is the angular response of the instrument that
in our case can be considered cilindrically simmetric (\ie
function of $\theta $ alone).

\subsection{Jupiter drifts.}

One of the most bright and point-like sub-millimetric sources
frequently adopted is Jupiter. The best way to determine the
beam-shape would be to drift-scan at various declination offsets
while chopping the source with a black-body reference, but
unfortunately atmospheric fluctuations (which would not be
removed), having a large ($\sim 16^{\prime}$) beam, dwarf the
diluted planet signal. Consequently we simulate the drift-scan
acquisition, introducing lock-in integration and correct
modulation by the acquired LVDT and pointing system. Each
beam-model generates a different convolved map from which the
drift is extracted. All drifts thus obtained are best-fitted to
the real time-ordered data (hereafter TOD) acquired. The advantage
of this simulation is the capacity of considering any degree of
sidelobe contamination during modulation if beamsize and half
beam-throw are comparable.

Resulting fits thus obtained define the best beam-shape for each
spectral channel. Solid angle $\Omega _{beam}$ present in
eq.\ref{form:resp} can then be computed by simple numerical
integration. Moreover, the different angular responses of the
various channels may induce differences in the ratios of
atmospheric fluctuations as they cross the observed sky patch,
differences that have to be accounted for during calibration. One
important issue is that the main beam and the reference beam do
not differ in the manner that contour levels of the beam are
similar and do not undergo deformations. This has been verified
confronting the simulated drift obtained by the afore-mentioned
procedure, and the following method to construct the reference
beam shape. This method is possible if data from a very small
parallactic angle $p$ is used. In our case 3 square-field
modulation allows us to reconstruct the reference beam by
rescaling the negative sidelobe at the beginning of the drift
scan. This can be done because contamination from the main beam is
lower than -50dB at distances greater than half beamthrow. A third
method to obtain the original beamshape is to generate families of
analytical beamshapes B(x) and simulate passage of the source by
calculating the contamination level for every data point as in
(eq.\ref{form:beam}):

\be\label{form:beam}
Data(x)=B(x)-\frac{1}{2}B(\sqrt{x^2+m_A^2-2xm_A\cos p}) \ee

Beam shapes obtained with different methods are consistent, and
FWHM obtained is equal to $16^{\prime}\pm 1^{\prime}$.

\begin{figure}[htbp]
    \begin{center}
       {\resizebox{!}{\textwidth}{\includegraphics*
            [28mm,80mm][200mm,260mm]{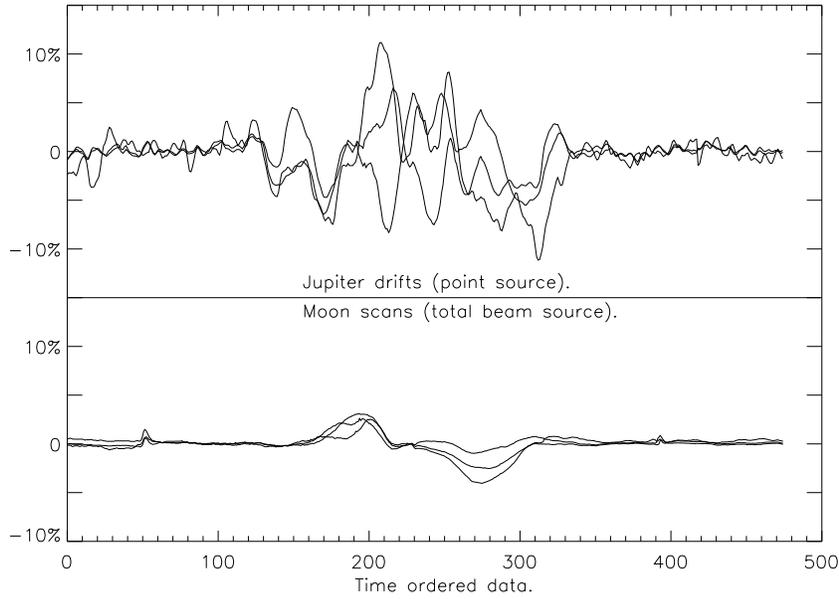}}}
        \caption{Relative differences of the 4 spectral channels
        over-plotted for Jupiter drifts (top) and moon scans (bottom). \label{fig:rapp}}
    \end{center}
\end{figure}

Moreover it is important to note that relative differences in the
beamshape, that can slightly alter the data on planet drifts, is a
negligible effect while studying beam-sized sources. Cross scans
performed on the moon with our instrument report very low relative
difference between different spectral channels (after intensity
peak-normalization). In fig. (\ref{fig:rapp}) relative differences
(i.e. difference of normalized scans - baseline $=0$, peak
emission $=1$ - between spectral channels) of Jupiter drifts and
Moon scans can be seen in the double plot. This allows us to
consider the same form-factor for all 4 channels in the analysis
of the galaxy cluster SZ-effect performed (De Petris et al. 2002).

\section{Refinement of optical calibrations.}

When detecting CMB anisotropies or indeed any cosmological signal
it is necessary to convert the voltage signal (amplified bolometer
output) to CMB thermodynamic temperature or power. To do so, a
good knowledge of the optical responsivity $ R=\Delta T/\Delta V$
is required. The best method to determine this physical parameter
would be of course, to measure the same signal that is our final
measuring objective but of which we already know its value (i.e.
CMB dipole for CMB experiments and so on...) This is not always
possible, so we restrain ourselves to dividing possible
calibrations in two different sets: point-like sources (total-flux
gathering), or extended sources (beam-dependent).

\subsection{Planet models.}\label{sec:planets}

\begin{figure}[htbp]
    \begin{center}
        {\resizebox{!}{\textwidth}{\includegraphics*
            [28mm,90mm][210mm,260mm]{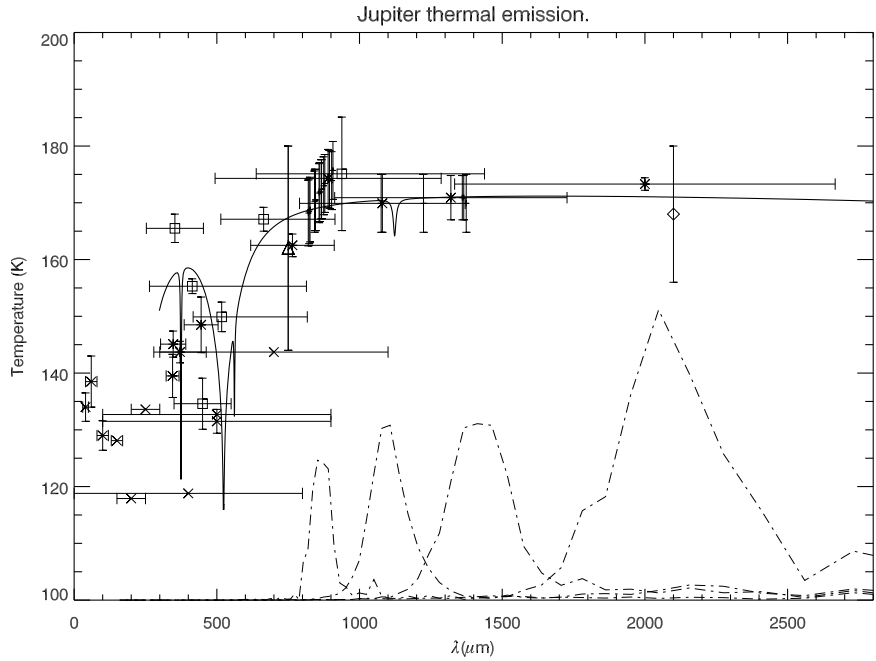}}}
        \caption{Jupiter temperature measures from
(see text references) different experiments. Over plotted is the
thermal emission with Van-Vleck modified profiles as in (R.Moreno
ph.D. thesis 1998). Over plotted are Fotomito spectral channels
over which integration is computed for calibration.}
 \label{fig:JupiterT}
    \end{center}
\end{figure}

Many models of planetary emission have been proposed and
subsequently refined with many different spectroscopic
measurements in the recent past (Ulich 1973), (Furniss et al.
1977), (Werner et al. 1978), (Whitcomb et al. 1978), (Hildebrand
et al. 1985), (Griffin et al. 1986), (Mangum 1992), (Naselsky et
al. 2003). All these measurements have been plotted in figure
fig.\ref{fig:JupiterT}, along with the thermal model with modified
Van-Vleck Weisskopf line profiles assumed by Moreno (Moreno,
1998). Moreover spectral absorption lines are expected from the
nominal synthetic spectra of Jupiter at the HCN (J=1 and J=2) and
PH3 (J=0) frequencies. The latter two should be practically
indistinguishable by a $\sim 10\%$ bandwidth filter as the ones we
employed. The variability of Jupiter's emission is known at low
frequencies to be connected to synchrotron emission relative to
spiraling particles of solar wind in Jupiter's magnetic field.
Although dominant at radio frequencies ($< 10 GHz$) this
contribution is negligible in the sub-mm region. The large
uncertainties in the existing Jupiter temperature measurements
hint to a possible apparent thermal variation, as noted in
(Naselsky et al. 2003), yet this could be the case if narrow
spectral bands were employed and local variations in chemical
composition distribution were present, altering the relative
absorption of the lines. In our case, integration over our
spectral channels, and spatially over the whole planet, would
dilute this variation at less than a few percent effect. Lack of
information, obtainable with future repeated measurements at
different time-scales with narrow-band spectrometers, could
resolve this uncertainty that leaves us with a $10\%$ error on the
thermal temperature of Jupiter used for calibrating our detectors.
These temperatures are for the 4 channels: 171, 171, 170 and 169K
respectively for the 143, 213, 272 and 353 GHz.

Substituting all quantities calculated and the total $\Omega$ of
the best-fitted beam in \ref{form:resp} we obtain the revised
responsivities for the four spectral channels : $(462 \pm 46)\mu
k/nV$, $(377 \pm 46)\mu k/nV$, $(426 \pm 43)\mu k/nV$ and $(317
\pm 32)\mu k/nV$. Ten percent errors have been considered due to
jupiter temperature uncertainty, widely dominant as error source.

\section{Geometrical form factors for cluster observations.}

\begin{figure}[htbp]
    \begin{center}
        {\resizebox{!}{\textwidth}{\includegraphics*
            [28mm,80mm][200mm,260mm]{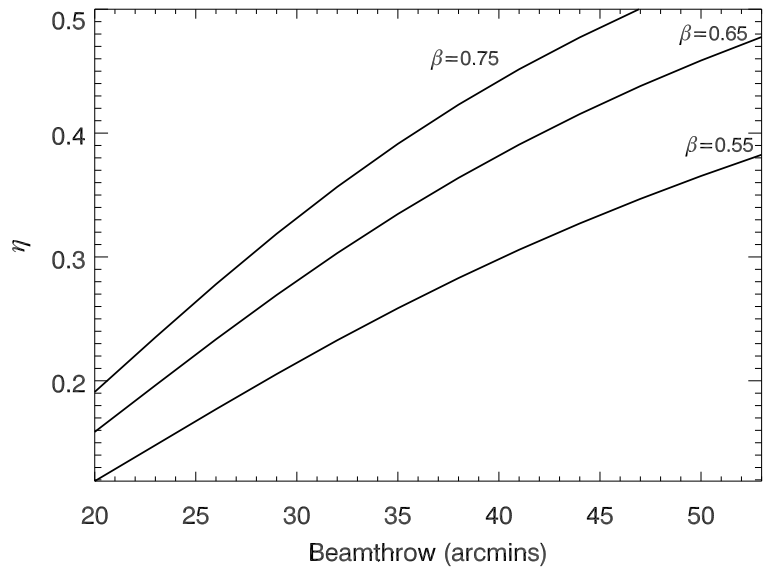}}}
        \caption{Form factor dependence from
beam-throw, when $3$-field modulation is applied. MITO telescope
  resolution ($16'$ FWHM) has been considered. A core radius of $8'$ has been assumed.\label{fig:ffvsbt}}
    \end{center}
\end{figure}

\begin{figure}[htbp]
    \begin{center}
        {\resizebox{!}{\textwidth}{\includegraphics*
            [28mm,80mm][200mm,260mm]{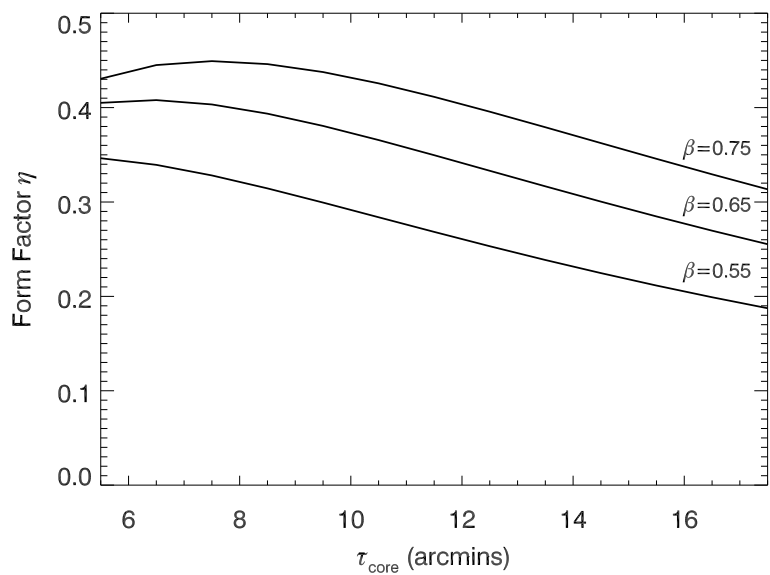}}}
        \caption{Form factor variation as a function
of core radius for the MITO telescope (FWHM $16'$), with the $\pm
20'$ $3$-field modulation pattern applied during the Coma
observations. A value of $\beta=0.75$ has been assumed, in
accordance with X-ray measurements on the same cluster. The
selected modulation amplitude ensures optimal efficiency
($\eta\simeq 0.4$) for $\theta_{core}\simeq 10'$ as for
Coma.\label{fig:ffvstcore}}
    \end{center}
\end{figure}

It has been mentioned before that the characterization of the beam
profile and determination of actual modulation pattern and
amplitude into the sky are vital to retrieve accurate calibration
data from the observation of point-like sources. Besides, they
also allow correct interpretation of observational data when
significantly extended sources are scanned; this is the case for
the measurement of SZ effect towards galaxy clusters (which
exhibit typical angular sizes of up to tens of arcminutes), where
signal recovery from noisy datasets can obviously take full
advantage from the characterization of filtering and selection
effects provided from the observation strategy. It has been noted
(see eq. \ref{form:resp}) that the effective brightness map
observed from the experiment when exploring a given sky region
will suffer from the dilution/smoothing effects due to the finite
angular resolution of the experiment:
\be
S_\nu(\alpha_0,\delta_0)=\int_{\alpha,\delta}B_\nu(\alpha-\alpha_0,\delta-\delta_0)AR(\alpha,\delta)d\Omega
\ee since, even at few tens of arcminutes away from the source
center, the collected signal will not be negligible, it turns out
that at a given modulation amplitude $\theta_{BT}$, a differential
measurement will be partially reduced from the contamination
effect in the side beams of the modulation pattern. It is
therefore necessary to distinguish between the {\it observed} SZ
effect, as the apparent CMB anisotropy that results from the
difference between beam-diluted signals in center and partially
contaminated reference beams at beam-throw $\theta_{BT}$, and the
{\it central} value of the comptonization across the cluster, as
it would be measured in absence of side beam contamination and in
pencil beam conditions. The two quantities, in term of the
measured CMB comptonization effect, are related by
\be
\Delta T^{obs}=\Delta T_{0}^{dil}-\Delta
T^{dil}_{\theta_{BT}}=\eta \Delta T_0 \ee where $\eta$ plays the
role of a geometrical {\it form factor} needed to define the
coupling between the instrument angular selection capabilities
(\ie  angular response and modulation beam-throw) and the cluster
morphology (\eg  the isothermal $\beta$-model parameters when such
modeling is justified from X-ray observation). Figs.
\ref{fig:ffvsbt} and \ref{fig:ffvstcore} show the dependence of
the form factor from beam-throw and cluster core radius
respectively, assuming the present single-pixel resolution of
$16'$ FWHM. Obviously, reference beam contamination decreases with
increasing beam-throws (fig.\ref{fig:ffvsbt}), and at any fixed
beam-throw the form factor is maximized for a certain specific
core radius representing a balance between reduction of signal
dilution and reference beam contamination, so that, basically, one
should try to reach the highest possible modulation amplitude for
any selected cluster.

More realistically, a compromise is to be found between
maximization of the form factor and removal of bulk atmospheric
fluctuation, which is more efficient when modulation explores sky
regions close to that of the main beam. The above considerations
have been usefully applied to evaluate $\eta$ in the case of the
Coma cluster, with the $\beta-$model parameters $b=0.75\pm 0.03$
and $\theta_c=10.5'\pm 0.2$, the beam profile reconstructed with
the procedure described in \ref{sec:beam} and a beamthrow of $\pm
18.5'$, we get $\eta=0.43\pm 0.02$ which is compatible, within
errors, with the determination of $0.41\pm 0.02$ reported in De
Petris \ea, 2002.

\section*{Acknowledgments} We wish to thank all the people that
worked at MITO project during all phases. Prof. P.Encrenaz and
T.Encrenaz for enlightening discussions on planet temperature
profiles, and dott. R.Moreno for providing the temperature profile
of Jupiter's emission. This work has been supported by CNR (Testa
Grigia Laboratory is a facility of Sezione IFSI in Turin),
COFIN-MIUR 1998, \& 2000, by ASI contract BAR.

\end{document}